# Magneto-transport and electronic structures in MoSi$_2$ bulks and thin films with different orientations

Wafa Afzal[1,3#], Frank Fei Yun[1,3#], Zhi Li[1,3], Zengji Yue[1,3*], Weiyao Zhao[1,3], Lina Sang[1,3], Guangsai Yang[1], Yahua He[1], Germanas Peleckis[1], Michael Fuhrer[2,3], and Xiaolin Wang[1,3*]

1. Institute of Superconducting and Electronic Materials (ISEM), Australian Institute for Innovative Materials (AIIM), University of Wollongong, Squires Way, North Wollongong, 2500, Australia.

2. School of Physics and Astronomy, Monash University, Clayton, Victoria, Australia Monash Centre for Atomically Thin Materials, Monash University, Clayton, Victoria, Australia

3. Future Low Energy Electronics Technologies (FLEET), Australian Research Centre (ARC), Australia.

[#] Equal contributions.

Email: zengji@uow.edu.au; xiaolin@uow.edu.au;

## Abstract

We report a comprehensive study of magneto-transport properties in MoSi$_2$ bulk and thin films. Textured MoSi$_2$ thin films of around 70 nm were deposited on silicon substrates with different orientations. Giant magnetoresistance of 1000% was observed in sintered bulk samples while MoSi$_2$ single crystals exhibit a magnetoresistance (MR) value of 800% at low temperatures. At the low temperatures, the MR of the textured thin films show weak anti-localization behaviour owing to the spin orbit coupling effects. Our first principle calculation show the presence of surface states in this material. The resistivity of all the MoSi$_2$ thin films is significantly low and nearly independent of the temperature, which is important for electronic devices.

## Introduction

Magnetoresistance (MR) has extremely wide applications ranging from magnetic sensors, magnetometers, electronic compasses, GPS navigation and magnetic storages (like magnetic cards and hard drives) [1-5]. Recently, a giant, non-saturating MR of ~ $10^7$% has been discovered in WTe$_2$ and MoSi$_2$ single crystals, which was attributed to electron-hole compensation and high mobilities of the carriers[6, 7]. Similar high MR has also recently been observed in many topological Dirac and Weyl semi-metallic systems such as Cd$_3$As$_{2}$[8], NbP[9], NbAs[10]. In many of these materials, the number of mobile holes is the same as the number of electrons[11]. For any material with anisotropic bands, MR depends on the orientation and the direction of current, because MR is governed by the Fermi contour determined by the current and magnetic field directions. The dependence of orientation and the direction of current has been observed on WTe$_2$, and provided essential information for understanding the band structure and topological physics of WTe$_2$. However, the MR dependence on orientation and direction of current of MoSi$_2$ has not been achieved yet.

In this work, we present a comprehensive study of surface termination and direction dependence of MR by fabricating and investigating different kinds of MoSi$_2$ samples, including single crystals, poly-crystals, textured thin films on Silicon substrates and the polycrystalline films on the oxide substrates. This gives an insight into the bulk and thin film characteristics of the material. The material in its bulk form shows a very high magnetoresistance which is modified when considered in the thin film form in certain orientations. A large MR of 1000% is observed in the polycrystalline sintered $MoSi_2$ bulk sample. The MR in $MoSi_2$ (110) thin film limit shows weak anti-localization (WAL) behaviour at low temperatures. First principle calculations including spin orbit coupling reveals spin-orbit interaction induced surface states, which is the origin of the observed WAL behaviour.

The main purpose of this work is 1) the fabrication of the $MoSi_2$ in thin films with different orientations; 2) the fabrication of single crystal and polycrystalline samples; 3) the study of magnetotransport of all these samples and compared to each other; 4) First principle calculation of electronic band structures of the $MoSi_2$ with and without doping; 5) understanding of possible mechanisms for the observed giant magnetoresistance in bulk samples and the absence of giant magnetoresistance but unusual magnetoresistance in thin film samples with different orientations. The purpose to investigate these kinds of samples is to probe the bulk as well as the thin film properties of the properties. As this material gives a very high MR (magnetoresistance) in its bulk forms and it is modified in the textured thin films. Behaviour of the MR for thin films has a striking dependence on the orientation of the films. The (110) oriented films show weak anti-localization behaviour which represents the presence of the surface states as also shown by the first principle calculations. This material seems to have a combination of trivial and topological states.

**Experiments and Results**

High purity $MoSi_2$ powder was pressed into pellets by using spark plasma technique. The Single crystal was grown in a floating zone furnace using $MoSi_2$ polycrystalline rods as the seed material. Textured thin films were deposited on silicon substrates with three different orientations by annealing the substrate pre-deposition[12], and polycrystalline films were grown on $LaAlO_3$ and $SrTiO_3$ substrates. At lower temperatures, hexagonal phase of the films is obtained, but annealing the films after deposition at temperatures around 700-800°C leads to a stable tetragonal phase.[13]

$MoSi_2$ films with (110) orientation were obtained on the silicon substrates with the orientation (111) and (100), and $MoSi_2$ films with (001) orientation were found on (101) Si substrates. The $MoSi_2$ films on oxide substrates $LaAlO_3$ and $SrTiO_3$ were found to be polycrystalline. The thickness of thin films was determined to be around 70 nm based on the x-ray reflectivity measurements. Fig.1 (a) shows the XRD scans of all the samples prepared in the study. The body centred tetragonal atomic structure of $MoSi_2$ is shown in Fig.1 (b-d). Atomic Force Microscopy (AFM) was carried out on $MoSi_2$ (001) films. Fig. 1(e) & 1(f) show the surface and phase scans for a selected area of scan of $1\mu m^2$. Smooth films with the surface roughness of less than a nanometre were observed. Scanning Electron Microscopy (SEM) was also carried out on a few samples. Figure 2(a) shows the cross sectional micrographs of the $MoSi_2$ (110) film on Si substrate. The line analysis shows the elemental distribution along the cross sectional direction which shows a little diffusion of Silicon at the interface between the substrate and the film. This means that the films are slightly Mo deficient at the interface but as we move towards the surface, the stoichiometry is restored. The Energy-dispersive X-ray spectroscopy (EDX) to show the chemical composition analysis of polycrystalline film on $LaAlO_3$ and textured film on Silicon is shown in Fig. 2(b, c). The samples were mechanically cut to ensure a smooth edge to be viewed. Magneto-transport measurements were carried out on $MoSi_2$ polycrystalline, single crystals, textured and polycrystalline thin films. In the transport measurements, we used standard four-probe configuration with the current applied in-plane.[14, 15] The measurements were conducted in a Physical Property Measurement System (PPMS, Quantum Design).

Firstly, the electrical resistivity of the samples is measured with the varying temperature in the absence of magnetic field using PPMS, and low resistivity values were obtained for all the samples as shown in Table 1. Fig. 3 (a) the resistivity in all the samples is a function of temperature. It was found that the resistivity of $MoSi_2$ thin films is higher than that in $MoSi_2$ bulks at the low temperature, but smaller at higher temperatures. Both $MoSi_2$ bulks and thin films show that the material behaves as a classical metallic conductor. A very small change (~ 3 μΩ) in the resistivity was observed in the $MoSi_2$ thin films over the temperature range of 3K to 300K, used in the experiments. The lowest resistivity was observed for the single crystal sample as expected owing to the lack of defects. The low electrical resistivity and the higher thermal stability of this material are the desired features in microelectronic devices. The largest change in resistivity (~ 120 μΩ) was observed for the pelletized sample. The trend of logarithmic increase in resistivity with the increase in temperature was observed in the bulk samples.

The MR was measured in all the samples, as displayed in Fig.3 (b-f). MR for single crystal, pellet and the (001) orientation films at higher temperatures have the parabolic dependence on the field. The linear fittings of MR versus $B^2$ are shown in the Fig. 4. The transverse MR for a material with two types of charge carriers with different mobilities is parabolic at lower fields as shown by our bulk samples. The highest value of MR was observed in the sintered sample pellet i.e. 1000 %. For single crystal, the value is lowered to 800%. The non-saturating behaviour of the MR in both the polycrystalline pellet and single crystal sample signals towards both electrons and holes taking part in conduction.

The MR behaviour of the films varies significantly from the crystal and pellet samples and their values are less than 5% for all films with different orientations. The MR curve of (001) orientation film shows almost linear behaviour in the low temperature regime and parabolic at higher temperatures. For the (110) orientation, however, there is a sharp cusp at the low field region at 3K and it saturates at high fields. For higher temperatures, MR drops significantly but shows a nearly linear behaviour. There are a few possible reasons for the big difference of MR behaviour between the bulk and thin films samples. The chemical compositions for the thin films should be slightly different from the crystal or pellet samples since films were fabricated and annealed at 700°C on silicon substrate, causing diffusions at the interface which is also evident through SEM measurements as shown in Fig. 2. If the material has surface states in addition to its bulk states due to SOC effect, the surface states should also contribute to the overall MR behaviour in thin film samples, while the bulk states would dominate in crystal and pellet samples. To find out if the SOC has strong effect on the band structures in $MoSi_2$, we performed calculations of band structure with and without SOC for both bulk and surface states.

The band structure in Fig.5 demonstrates the co-existence of holes (Γ-point) and electrons (Z point) in the bulk Brillouin zone. The majority charge carrier contributions are from the *d* and *p* orbitals of the Mo atoms at the fermi level as shown in Fig. 5(b). From Fig. 5a, if we zoom in to the energy range (-0.8 to 0.2) as shown in Fig. 5c, we find that the bands at Z point are clearly split by the SOC and the gap is significantly widened for both upper and lower bands. For the upper bands, the gap is nearly 0.1 eV and the lower band has a gap of ~0.05eV. This means that the spin orbit interaction (SOI) in this compound is indeed strong. The bands crossing the Fermi level do not show much difference with the SOI, which indicates that the SOI should have a weak effect on the bulk transport properties. Due to the large SOI strength, we performed calculations of surface bands, some of which are shown in Fig. 6. The detailed calculation results of the surface band structure are included in the supporting information.

The linear dependence of transverse MR on the magnetic field, as shown by MoSi$_2$ (001) film, has been discussed in different theories and models. It can be attributed to either quantum or classical effects. According to the quantum version of the linear MR (LMR) it originates from the linear bands or the occupation of the lowest Landau level under the application of field as explained by Abrikosov[16], this should yield high value for MR. As our first principle calculations show the parabolic band structure, this quantum MR cannot be the reason for the linear MR in our samples. Many classical models and studies on the linear MR hold the presence of defects and the mobility fluctuation of the carriers responsible for this linear behaviour as explained by the Parish Littlewood Theory.[17-19] Another 3D model for metal-semiconductor composites explains LMR originating from the Gaussian distribution of the mobility in the 3D samples[20]. The linear behaviour of the MR at lower temperatures in MoSi$_2$ (001) indicates disorder driven behaviour.

Along (110) orientation of the MoSi$_2$ films, however, the MR shows a cusp at the lower fields at lower temperatures which can be attributed to the weak anti localization effects[21, 22] also observed in other topological thin film systems such as Cd$_3$As$_2$[23], Bismuth Selenide[24], Bi$_2$Se$_2$Te[25], DyPdBi[26]. It is an interesting finding as this quantum effect is observed due to strong intrinsic spin orbit coupling[27-29], the extrinsic impurity spin orbit effects[30] or the topological nature of the surface states[31]. In the presence of spin orbit coupling, the quantum correction to the conductivity can be explained using HLN (Hikami-Larkin-Nagaoka) equation. At low temperatures, diffusive transport systems with little disorder can be modelled using the HLN equation[21, 32]

$$\Delta \sigma_{HLN}(B) = \alpha \frac{e^2}{\pi h} \left[ \psi \left( \frac{\hbar}{4eB\mathcal{L}^2} + \frac{1}{2} \right) - \ln \left( \frac{\hbar}{4eB\mathcal{L}^2} \right) \right]$$

Where $\Delta\sigma(B)$ represents change in magneto-conductivity, $\psi$ is the digamma function, $e$ is the electronic charge, $h$ is the Planck's constant, $\frac{\hbar}{4e\mathcal{L}^2}$ is the characteristic magnetic field, B is the applied magnetic field, $\mathcal{L}$ is the phase coherence length and $\alpha$ is a pre-factor that indicates the type of localization and the number of surface bands.

For our results, the fittings were made using the HLN equation on the thin film samples, the fitted curve is shown in Fig.7. The fitted phase coherence length for MoSi$_2$ (110)/Si (111) is 44 nm with α pre-factor of -2.2 "α" takes on a value based on the number of conducting channels in the system. For a single conducting channel, it takes the value of 0.5, however, the mixing of states can lead to a smaller value. The WAL behaviour in our thin film samples point towards the strong spin orbit contribution in the transport. The anisotropy of the electronic transport along different orientations of the structure becomes evident from the observation of the weak anti-localization effects along this orientation and linear behaviour even at higher fields on the other. This change of behaviour along different directions also implies presence of both protected and conventional states in this material which has been pointed out previously[6].

For the polycrystalline films obtained on two oxide substrates, the MR is nearly zero. We believe that this decrease in the MR is caused by the metal diffusion between the substrate and the film that is evident from the EDX spectra obtained through SEM for the film deposited on LAO substrate. Metals follows a nearly free electron model, in which the high MR effects are not observed if the electrons and hole densities are different. To confirm this, the density of states (DOS) was calculated for Aluminium introduced MoSi$_2$ and it shows that the DOS is sufficiently increased by this metallic diffusion as shown in the Figure 8. Calculation of Al diffusion in MoSi$_2$ were performed by substitutional doping of Mo with Al at one Mo site for every 2x2 supercell of MoSi$_2$ resulting in an atomic percentage of 2.1 at. %. The 2x2 structure was then optimized until 10$^{-6}$ eV/Å. A 36×36×16 K-point mesh was used to sample the Brillouin zone with a cut off energy of 500eV for the plane wave basis.

The Hall measurements were carried out on the MoSi$_2$ (110)/Si (111) thin film samples, as shown in Fig.9 (a). In the previous report for MoSi$_2$ single crystals, both the electrons and holes are deemed responsible for the transport. The Hall resistance for our measured samples do not show sign reversal with the change in temperature and the negative slope indicates that the charge carriers are electrons. The mobilities calculated by the Hall measurements for the thin film samples are low (~18cm$^2$/Vs) and the Hall behaviour is nearly independent of the temperature.

For the materials following a multi band transport, with quadratic MR, the mobility is calculated by the MR data using the formula $\mu = \frac{(\Delta R/R)^{1/2}}{B}$ [33]. To gain an insight into the transport properties, it is essential to consider both MR and Hall measurements. Fig. 9 (b) shows a plot of these mobilities against temperature. The values for the pellet and the single crystal show a similar trend and the highest mobility is exhibited by the pellet sample (3162cm$^2$/Vs). Correlating this behaviour with the behaviour of change in resistance with the temperature, it can be deduced that as there is not much change in the measured resistivity of the films with temperature, they can be better understood by the Hall measurements i.e. the representative mobility for thin film samples is the one calculated by the Hall measurements. For the bulk samples however, the change in the resistivity with temperature is quite defined as shown in Fig.3 (a), also represented by the MR curves, so the representative mobility would be MR mobility in that case with contributions from both types of carriers as the calculated band structure also depicts.

**Analysis and discussion**

There is a metallic behaviour observed from the electrical resistivity for all the samples. For the case of bulk samples, this behaviour is quite pronounced as shown by the measurements. The thin films show the similar behaviour with different values. The single crystal, the ingot and the polycrystal show a lower resistivity as compared to the thin films. As mentioned before, we have observed the diffusion effects and the decreased mobility in the thin films. This causes the resistivity to increase as compared to its bulk counterparts. Any kind of impurity factors in the films can lead to a resistivity that does not vary much with the temperature.

From the results, we can see that the electronic properties of MoSi$_2$ depend on their composite formats, substrates and growth orientation. The MoSi$_2$ bulk samples have higher mobility and MR values while the MoSi$_2$ thin films have lower mobility and MR values. The MR also has different dependence relations with the temperature which show the phonon contributions becoming relevant at higher temperatures and decreasing the MR. The MR in MoSi$_2$ bulk systems is parabolic and non-saturating at high magnetic fields while for the thin films, it shows different behaviour for different orientations. For (001) oriented films, it is linear at lower temperatures pointing towards the disorder effects. MoSi$_2$ (110) oriented films, however, show weak anti-localization behaviour driven by the SOC in this material.

MoSi$_2$ demonstrates tunable MR property in both bulk and thin film samples depending upon the interface effects between MoSi$_2$ films and substrates. The thin films on Silicon are oriented with uniform surfaces as compared to the ones deposited on the oxide substrates due to the strain between MoSi$_2$ films and substrates, which can modify the electronic structure.[34, 35]. Therefore, the electronic properties in MoSi$_2$ can be modified through interface engineering.

This unique property could find important applications in tunable magnetic devices like magnetic sensors. The resistivity in MoSi$_2$ films with (110) orientations grown on Si substrates is as low as $5\times10^{-5}$ $\Omega$cm and is nearly independent of temperature. This is an excellent electrical property and could find wide applications in low-consumption electronic devices. In modern electronic devices, the performance and capability starts to decrease with increasing temperatures. The increased temperature may come from waste heat or current caused thermal effects. Through improving electron mobility, by either gating or doping, MoSi$_2$ films could be used for designing and fabricating nano-electronic devices.

Several concepts were systematically investigated in our first principle calculations. Firstly, to find out if the SOC has a strong effects on the band structures in MoSi$_2$, we performed calculations of band structure with and without SOC for both bulk and surface states. We examined the strength of the SOC through the use of the E-K relationship and determined that the system has a SOC strength of 50-100 meV, which indicated that it indeed has a substantial SOC. Secondly, by looking into the partial densities of states in our first principle calculation we were able to determine that the main charge carriers of the systems are the *d* orbitals of the Mo ions and p orbitals of the Si ions. Thirdly, we looked at whether the strong SOC would have any effect on the surface states of the crystals and found that surface state exists in the crystal, which is comparable to the fitted surface states found in our WAL Magnetoresistance measurements. Lastly, we looked at the surfaces states of the and found that the crystal is topologically non-trivial, with a $\mathcal{Z}_2$ index $(v_0;1v_2v_3)$ $of$ (0;111) which indicates that it is a weak topological insulator.

**Conclusions**

We conducted comprehensive studies on the synthesis, characterization and magnetotransport measurements of MoSi$_2$ bulk and thin film samples with different orientations. The highest MR value is exhibited by the sintered samples of polycrystalline MoSi$_2$. The MoSi$_2$ single crystal doesn't display the giant MR as reported earlier, which can be attributed to the possible lack of stoichiometry in our single crystal sample. The MR in MoSi$_2$ thin film samples is different from the bulk and shows different character in different orientations. MoSi$_2$ (001) orientation shows impurity driven linear behaviour while (110) orientation shows SOC induced weak anti-localization effects. A detailed DFT calculation identified spin-momentum locked surface states.

**Methods**

Three types of samples were prepared to make a qualitative comparison between the changes in the transport properties. Single crystal was fabricated using pure polycrystalline MoSi$_2$ cylindrical rods as the seed material for crystal growth in the floating zone furnace. A small crystal of around 8mm in length was obtained. Small flakes of the crystal(~3x1.5mm) were used for the transport measurements by attaching gold wires to make connections with the PPMS measurement puck. Sintered pellets were made using MoSi$_2$ powder using Spark Plasma Sintering. The pressure applied to the sample was 50 MPa at the temperature of 1200 $^o$C. Disc shaped pellets of around 1.5cm radius were obtained. A few parts were cut and polished in small rectangular pieces of 4x3mm for the magneto-transport measurements. MoSi$_2$ thin films were deposited using pulsed laser deposition (PLD) technique. MoSi$_2$ polycrystalline target was used for the deposition of films. Insulating substrates of Si were cut into small rectangular pieces (~5x4mm) and were ultrasonically cleaned for ten minutes and then dried. Deposition was carried out on different orientations of Silicon substrates and two oxide substrates LaAlO$_3$ and SrTiO$_3$. All the substrates were annealed before deposition at 700 $^o$C-750 $^o$C in vacuum close to $4\times10^{-4}$ Pa. The depositions were made at the temperature of 500 °C at the vacuum level of $3\times10^{-4}$ Pa for 45 minutes. The laser power during the deposition process was set to 2W.

The transport measurements were done on a Quantum Design 14 T Physical Property Measurement System (QD-PPMS). Gold contacts were deposited on the film using a sputter coater. Gold wires were then used to make the contacts using silver paste. Atomic Force Microscopy (AFM) was carried out on a few samples. Model MFP-3D from Asylum Research was used in non-contact mode. The sample was fixed on a slide using double-side adhesive tape. Scanning electron Microscopy (SEM) was also done on a few samples to observe the interface between the film and the substrate and Energy dispersive X-ray (EDX) analysis. The thickness of the films was found by the using X-ray reflectometry.

First principle calculations were performed using the density function theory (DFT) implemented by the Vienna Ab initio Simulation Package (VASP). The exchange-correlation function used is the General Gradient Approximation (GGA) with Perdew-Burke-Ernzerhof (PBE) formulation. Structural optimization was performed, and the atomic position and cell vectors were relaxed until the energy, maximum force, and maximum displacement were less than $5\times10^{-6}$ eV and 0.01 eV/A, a $36\times36\times16$ K-point mesh was used to sample the Brillouin zone with a cut off energy of 500eV for the plane wave basis.

## Acknowledgments

This research is supported and funded by ARC Centre for Future Low Energy Electronics Technologies (ARC-FLEET), ARC Professional Future Fellowship (FT130100778).
We would also like to thank Dr. Sheikh Muhammad Kazi Nazrul Islam of our group, Dr. Tony Romeo and Dr. Mitchell Nancarrow at Electron Microscopy Centre, AIIM, and UoW for their help in microscopic measurements.

**Supporting Information available:** This includes the calculated surface electronic structure, Fermi surfaces and spin textures along them.

## References

[1] A.M. Deac, A. Fukushima, H. Kubota, H. Maehara, Y. Suzuki, S. Yuasa, Y. Nagamine, K. Tsunekawa, D.D. Djayaprawira, N. Watanabe, Bias-driven high-power microwave emission from MgO-based tunnel magnetoresistance devices, Nature Physics, 4 (2008) 803.
[2] S.X. Wang, G. Li, Advances in Giant Magnetoresistance Biosensors With Magnetic Nanoparticle Tags: Review and Outlook, IEEE Transactions on Magnetics, 44 (2008) 1687-1702.
[3] J. Slaughter, Materials for magnetoresistive random access memory, Annual Review of Materials Research, 39 (2009) 277-296.
[4] D. Rifai, A.N. Abdalla, K. Ali, R. Razali, Giant Magnetoresistance Sensors: A Review on Structures and Non-Destructive Eddy Current Testing Applications, Sensors (Basel, Switzerland), 16 (2016) 298-298.
[5] Z.J. Yue, X.L. Wang, S.X. Dou, Angular-dependences of giant in-plane and interlayer magnetoresistances in Bi2Te3 bulk single crystals, Applied Physics Letters, 101 (2012) 152107.
[6] M. Matin, R. Mondal, N. Barman, A. Thamizhavel, S.K. Dhar, Extremely large magnetoresistance induced by Zeeman effect-driven electron-hole compensation and topological protection in ${\mathrm{MoSi}}_{2}$, Physical Review B, 97 (2018) 205130.
[7] M.N. Ali, J. Xiong, S. Flynn, J. Tao, Q.D. Gibson, L.M. Schoop, T. Liang, N. Haldolaarachchige, M. Hirschberger, N.P. Ong, R.J. Cava, Large, non-saturating magnetoresistance in WTe2, Nature, 514 (2014) 205.
[8] T. Liang, Q. Gibson, M.N. Ali, M. Liu, R.J. Cava, N.P. Ong, Ultrahigh mobility and giant magnetoresistance in the Dirac semimetal Cd3As2, Nature Materials, 14 (2015) 280-284.

[9] C. Shekhar, A.K. Nayak, Y. Sun, M. Schmidt, M. Nicklas, I. Leermakers, U. Zeitler, Y. Skourski, J. Wosnitza, Z. Liu, Y. Chen, W. Schnelle, H. Borrmann, Y. Grin, C. Felser, B. Yan, Extremely large magnetoresistance and ultrahigh mobility in the topological Weyl semimetal candidate NbP, Nature Physics, 11 (2015) 645-649.
[10] N.J. Ghimire, Y. Luo, M. Neupane, D.J. Williams, E.D. Bauer, F. Ronning, Magnetotransport of single crystalline NbAs, Journal of Physics: Condensed Matter, 27 (2015) 152201.
[11] I. Pletikosić, M.N. Ali, A.V. Fedorov, R.J. Cava, T. Valla, Electronic Structure Basis for the Extraordinary Magnetoresistance in ${\mathrm{WTe}}_{2}$, Physical Review Letters, 113 (2014) 216601.
[12] A. Perio, J. Torres, G. Bomchil, F. Arnaud d'Avitaya, R. Pantel, Growth of MoSi2 with preferential orientation on (100) silicon, Applied Physics Letters, 45 (1984) 857-859.
[13] J.W.C. de Vries, A.H. van Ommen, Transport properties of hexagonal and tetragonal MoSi2 thin films, Journal of Applied Physics, 64 (1988) 749-752.
[14] Z. Yue, I. Levchenko, S. Kumar, D. Seo, X. Wang, S. Dou, K. Ostrikov, Large networks of vertical multi-layer graphenes with morphology-tunable magnetoresistance, Nanoscale, 5 (2013) 9283-9288.
[15] Z.J. Yue, X.L. Wang, S.S. Yan, Semimetal-semiconductor transition and giant linear magnetoresistances in three-dimensional Dirac semimetal Bi0.96Sb0.04 single crystals, Applied Physics Letters, 107 (2015) 112101.
[16] A.A. Abrikosov, Quantum magnetoresistance, Physical Review B, 58 (1998) 2788-2794.
[17] M.M. Parish, P.B. Littlewood, Classical magnetotransport of inhomogeneous conductors, Physical Review B, 72 (2005) 094417.
[18] P. Li, Q. Zhang, X. He, W. Ren, H.-M. Cheng, X.-x. Zhang, Spatial mobility fluctuation induced giant linear magnetoresistance in multilayered graphene foam, Physical Review B, 94 (2016) 045402.
[19] J. Hu, M.M. Parish, T.F. Rosenbaum, Nonsaturating magnetoresistance of inhomogeneous conductors: Comparison of experiment and simulation, Physical Review B, 75 (2007) 214203.
[20] J. Xu, D. Zhang, F. Yang, Z. Li, Y. Pan, A three-dimensional resistor network model for the linear magnetoresistance of Ag2+δSe and Ag2+δTe bulks, Journal of Applied Physics, 104 (2008) 113922.
[21] H.-Z. Lu, S.-Q. Shen, Weak localization of bulk channels in topological insulator thin films, Physical Review B, 84 (2011) 125138.
[22] Z. Yue, K. Rule, Z. Li, W. Zhao, L. Sang, G. Yang, C. Tan, L. Wang, A. Bake, X. Wang, Weak localization and anti-localization in rare earth doped topological insulators, arXiv preprint arXiv:2008.03919, (2020).
[23] B. Zhao, P. Cheng, H. Pan, S. Zhang, B. Wang, G. Wang, F. Xiu, F. Song, Weak antilocalization in Cd3As2 thin films, Scientific Reports, 6 (2016) 22377.
[24] P. Sahu, J.-Y. Chen, J.C. Myers, J.-P. Wang, Weak antilocalization and low-temperature characterization of sputtered polycrystalline bismuth selenide, Applied Physics Letters, 112 (2018) 122402.
[25] R.K. Gopal, S. Singh, R. Chandra, C. Mitra, Weak-antilocalization and surface dominated transport in topological insulator Bi2Se2Te, AIP Advances, 5 (2015) 047111.
[26] V. Bhardwaj, S.P. Pal, L.K. Varga, M. Tomar, V. Gupta, R. Chatterjee, Weak Antilocalization and Quantum Oscillations of Surface States in Topologically Nontrivial DyPdBi(110)Half Heusler alloy, Scientific Reports, 8 (2018) 9931.
[27] Y. Li, Z. Wang, P. Li, X. Yang, Z. Shen, F. Sheng, X. Li, Y. Lu, Y. Zheng, Z.-A. Xu, Negative magnetoresistance in Weyl semimetals NbAs and NbP: Intrinsic chiral anomaly and extrinsic effects, Frontiers of Physics, 12 (2017) 127205.
[28] H. Nakamura, D. Huang, J. Merz, E. Khalaf, P. Ostrovsky, A. Yaresko, D. Samal, H. Takagi, Robust weak antilocalization due to spin-orbital entanglement in Dirac material Sr3SnO, Nature Communications, 11 (2020) 1161.
[29] X. Huang, L. Zhao, Y. Long, P. Wang, D. Chen, Z. Yang, H. Liang, M. Xue, H. Weng, Z. Fang, X. Dai, G. Chen, Observation of the Chiral-Anomaly-Induced Negative Magnetoresistance in 3D Weyl Semimetal TaAs, Physical Review X, 5 (2015) 031023.

[30] W.E. Liu, E.M. Hankiewicz, D. Culcer, Quantum transport in Weyl semimetal thin films in the presence of spin-orbit coupled impurities, Physical Review B, 96 (2017) 045307.
[31] C.-Z. Chang, P. Tang, Y.-L. Wang, X. Feng, K. Li, Z. Zhang, Y. Wang, L.-L. Wang, X. Chen, C. Liu, W. Duan, K. He, X.-C. Ma, Q.-K. Xue, Chemical-Potential-Dependent Gap Opening at the Dirac Surface States of $\mathrm{Bi}_2\mathrm{Se}_3$ Induced by Aggregated Substitutional Cr Atoms, Physical Review Letters, 112 (2014) 056801.
[32] S. Hikami, A.I. Larkin, Y. Nagaoka, Spin-Orbit Interaction and Magnetoresistance in the Two Dimensional Random System, Progress of Theoretical Physics, 63 (1980) 707-710.
[33] T.L. Martin, J.E. Mahan, Electronic transport and microstructure in MoSi2 thin films, Journal of Materials Research, 1 (1986) 493-502.
[34] S.J. Callori, S. Hu, J. Bertinshaw, Z.J. Yue, S. Danilkin, X.L. Wang, V. Nagarajan, F. Klose, J. Seidel, C. Ulrich, Strain-induced magnetic phase transition in $\mathrm{SrCoO}_{3-\delta}$ thin films, Physical Review B, 91 (2015) 140405.
[35] S. Hu, Z. Yue, J.S. Lim, S.J. Callori, J. Bertinshaw, A. Ikeda-Ohno, T. Ohkochi, C.-H. Yang, X. Wang, C. Ulrich, J. Seidel, Growth and Properties of Fully Strained SrCoOx (x ≈ 2.8) Thin Films on DyScO3, Advanced Materials Interfaces, 2 (2015) 1500012.

**Figures and Figure Captions**

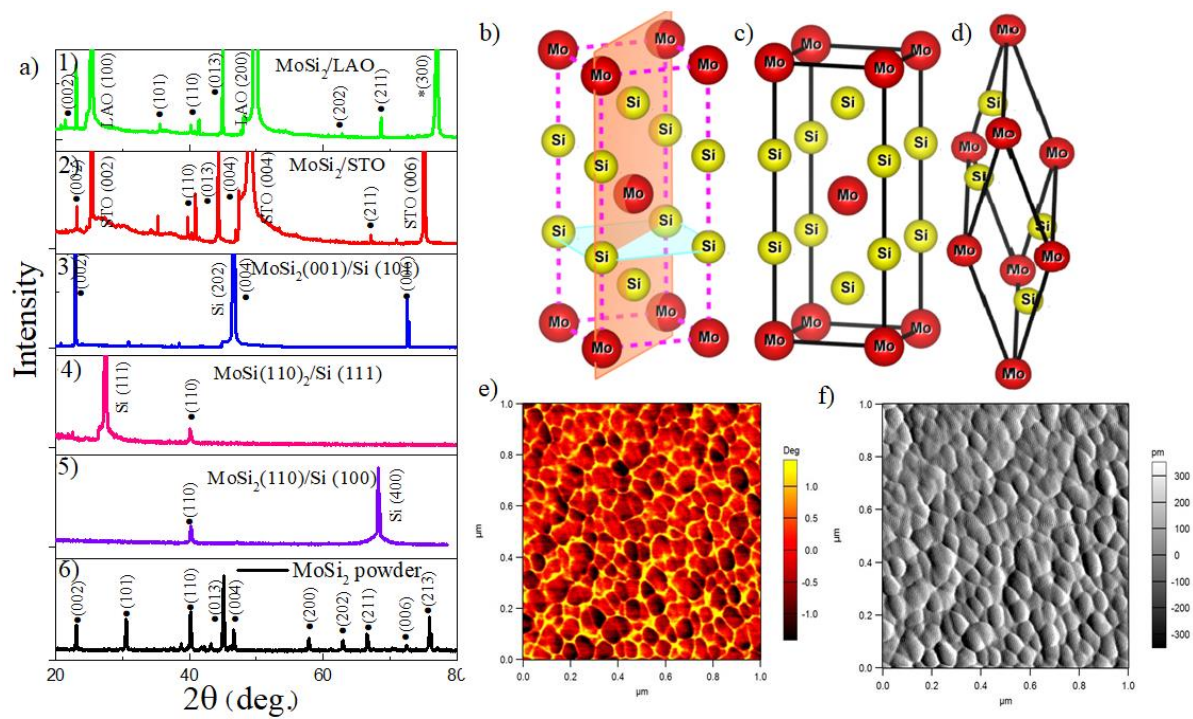

**Figure 1**. **a)** X-Ray Diffraction images of 1) MoSi$_2$ film on (100) LaAlO$_3$. 2) MoSi$_2$ film on (001) SrTiO$_3$. 3) (001) MoSi$_2$ film on Si (101). 4) (110) MoSi$_2$ film on Si (111). 5) (110) MoSi$_2$ film on Si (100). 6) MoSi$_2$ ingot. The peaks for polycrystalline ingot show a pure and clean phase of MoSi$_2$. Si (100) and Si (111) favour the growth of (110) direction of MoSi$_2$. The oxide substrates, however, show a polycrystalline, but a single phase deposition. **b)** The unit cell for body centred tetragonal crystal structure of MoSi$_2$. **c)** Primitive cell for MoSi$_2$ crystal structure. **d)** Unit cell showing (110) and (001) planes of MoSi$_2$. **e, f)** AFM phase and height scans of MoSi$_2$ (001) orientation thin film of 1μm$^2$ area.

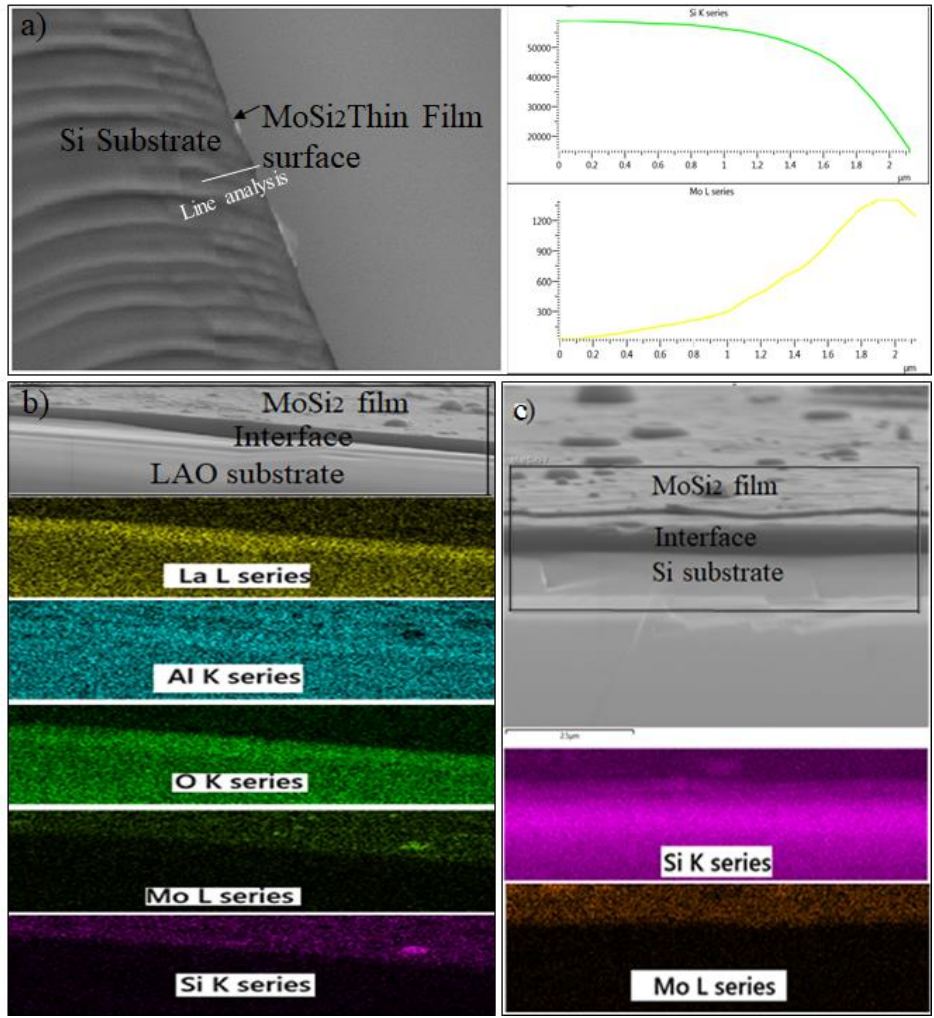

**Figure 2. a)** Cross sectional SEM image and line analysis of MoSi$_2$ (110) film on Silicon substrate showing the element distribution w.r.t distance. Cross sectional SEM images and EDX analysis of **b)** Polycrystalline MoSi$_2$ film on LAO substrate **c)** (110) MoSi$_2$ film on Si substrate.

**Table 1**: Comparison of resistivity values of different samples at 3K and 300K.

|  | Sintered Pellet | Ingot | Single crystal | MoSi$_2$(110) on Si(100) | MoSi$_2$(110) on Si(111) | MoSi$_2$(001) on Si(101) |
|---|---|---|---|---|---|---|
| ρ at 3K | ~11 μΩcm | 3 μΩcm | ~2 μΩcm | 49 μΩcm | 59 μΩcm | 49 μΩcm |
| ρ at 300K | 131 μΩcm | 47 μΩcm | 18 μΩcm | 51 μΩcm | 62 μΩcm | 51 μΩcm |

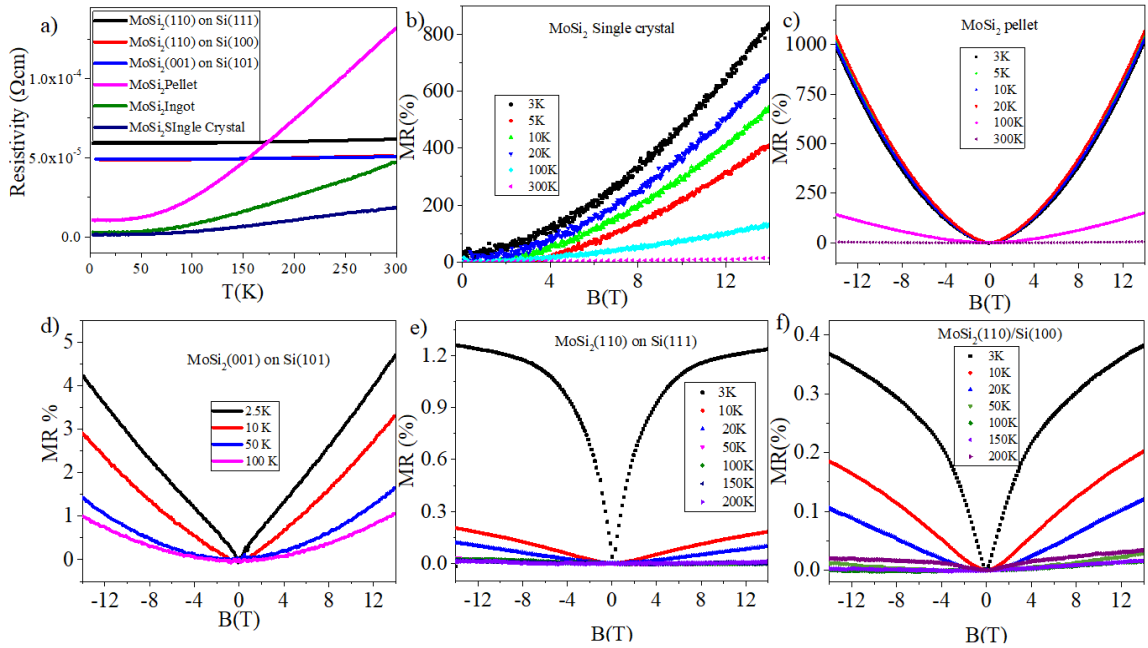

**Figure 3.a)** Resistivity vs. temperature measurements for MoSi$_2$ (110) and (001) thin films, pellet, ingot and single crystal samples. The resistivity of thin films shows a behaviour different from the bulk counterparts, with a little variation in the resistivity with increasing temperature. Magnetoresistance vs B plots of **b)** Single crystal **c)** Sintered pellet **d)** (001) MoSi$_2$ film on Si (101) **e)** (110) MoSi$_2$ film on Si (111) **f)** (110) MoSi$_2$ film on Si (100).

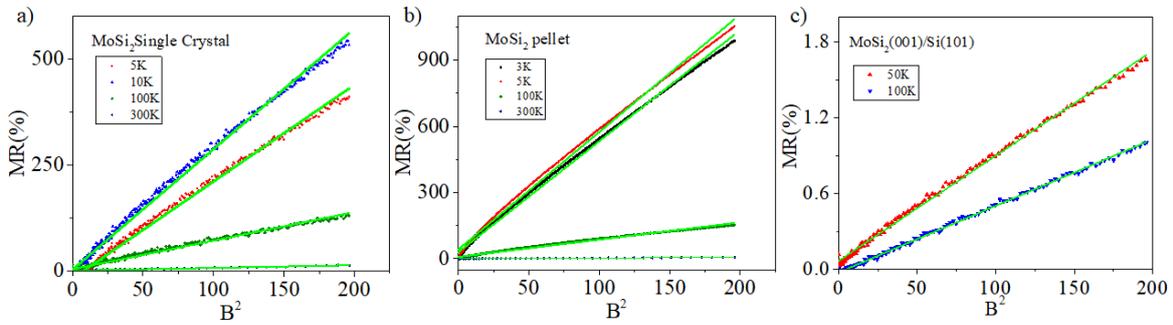

**Figure 4**: MR vs B$^2$ plots and the linear fits (green) for MoSi$_2$ **a)** Single crystal sample **b)** Pellet and **c)** (001) oriented film on Si (101) substrate at higher temperatures, showing the quadratic nature of MR.

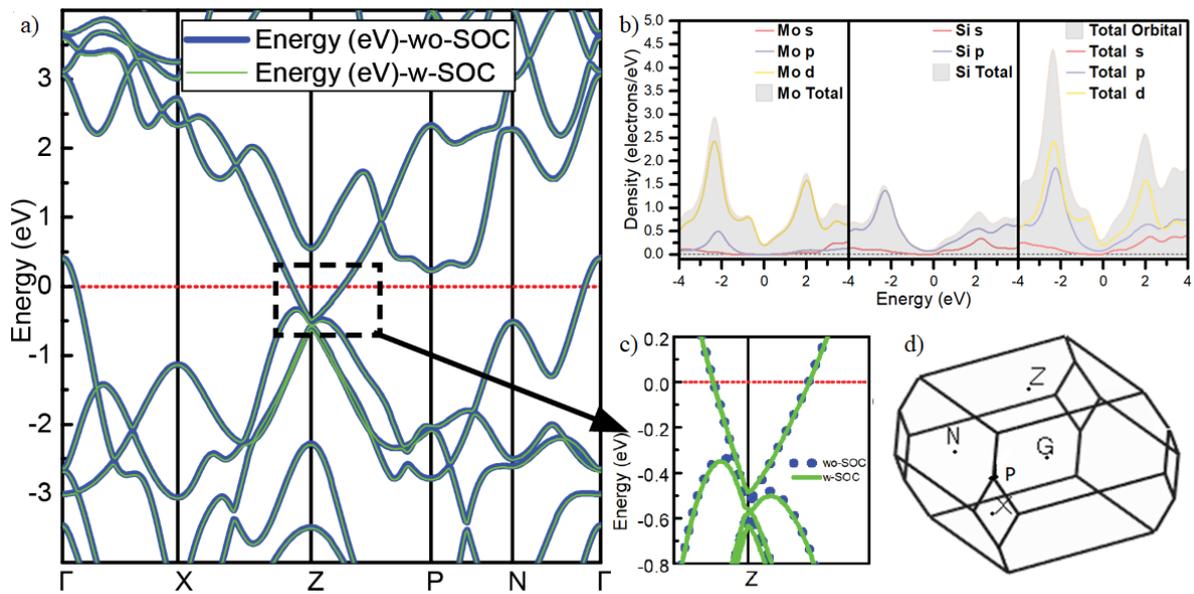

**Figure 5**: The Electronic structures of MoSi$_2$. **a)** Band structures with (w) and without (wo) spin orbit coupling **b)** The density of states profile for Mo and Si atoms at different energies **c)** Zoomed in band structure in a low energy range showing gap opening at Z point with the inclusion of SOC **d)** Brillouin Zone for MoSi$_2$ showing high symmetry points.

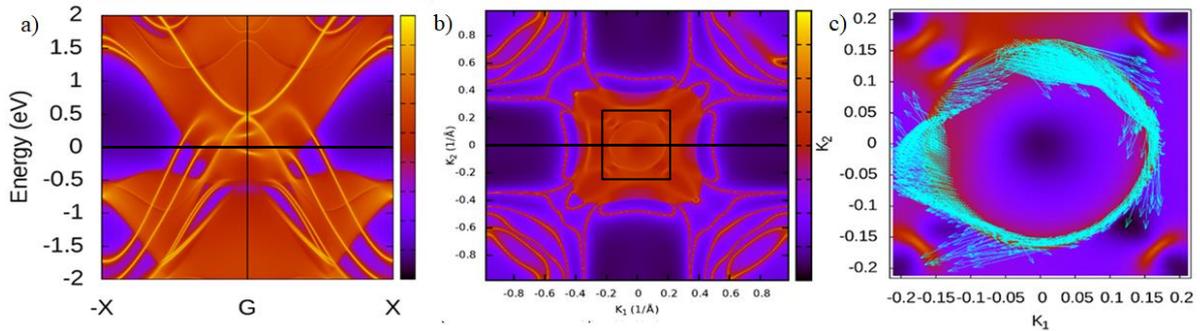

**Figure 6**: First Principle calculation for the surface states in MoSi$_2$ showing **a)** The surface bands along $X' - \Gamma - X$ direction highlighting the surface bands (yellow) in the first Brillouin zone **b)** Fermi surface and **c)** Spin textures showing spin momentum locking along fermi surfaces.

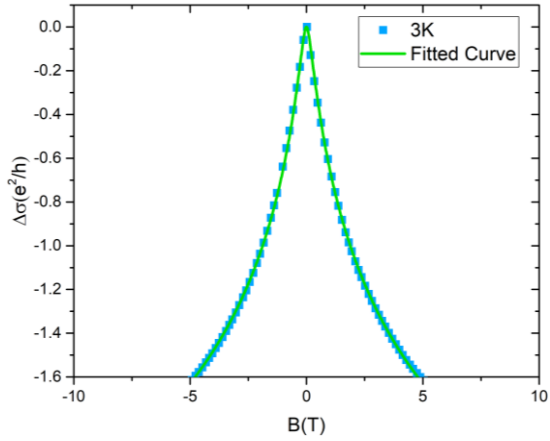

**Figure 7**: HLN equation fitting (green) on the magneto-conductivity (blue) of the $MoSi_2$ (110) film on Si (111) substrate at 3K.

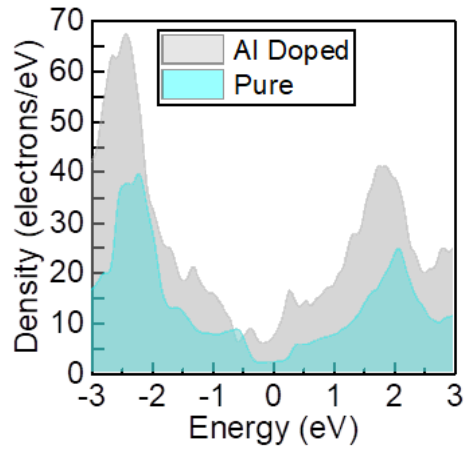

**Figure 8**: The comparison of density of states (DOS) of pure and Aluminium doped $MoSi_2$.

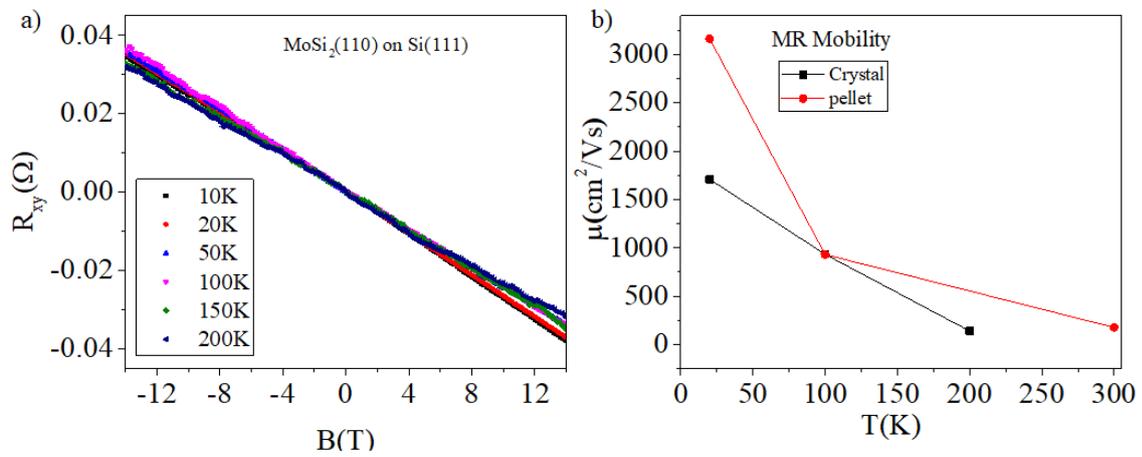

**Figure 9: a)** The Hall resistance graph for the (110) oriented MoSi$_2$ thin film on Si substrate. The negative slope shows that electrons are the charge carriers. **b)** The MR mobility calculated for Pellet and Single crystal samples using the formula $\mu = \frac{(\Delta R/R)^{1/2}}{B}$ .